\documentclass{aa}  
\usepackage{ragged2e}
\usepackage{inconsolata}
\usepackage{float}
\usepackage{graphicx}
\usepackage{tablefootnote}
\usepackage{academicons}
\usepackage{tikz,xcolor,hyperref}
\newcommand{\orcidauthorA}{0009-0009-8988-0537}
\newcommand{\orcidauthorB}{0000-0001-6264-140X}
\newcommand{\orcidauthorC}{0000-0003-0079-1239}
\newcommand{\orcidauthorD}{0000-0002-0273-218X}
\usepackage{orcidlink}
\hypersetup{
    colorlinks=true,
    citecolor=blue,
    linkcolor=blue,
    urlcolor=blue,
}
\usepackage{natbib}
\usepackage{txfonts}
\usepackage{comment}

\def\medd{$\dot{m}_{\rm Edd}$}

\def\swift{{\it Swift}}

\def\spin{$\alpha^{*}$}
\def\ltransf{$L_{\rm transf}/L_{\rm disc}$}
\def\fcol{$f_{\rm col}$}

\def\psilt{$\Psi_{\rm \lambda}(t)$}
\def\fobst{$F^{\rm obs}_{2-10}(t)$}

\begin{document} 


\title{The X-ray/UV Connection in NGC 5548: A Rapidly Varying Corona}

\author{M. Papoutsis \inst{1,2}
\and
I. E. Papadakis\inst{3,2}
\and
C. Panagiotou\inst{4} 
\and 
E. Kammoun\inst{5}
\and
M. Dovčiak\inst{6}
}

\titlerunning{ngc5548 light curve fitting}
\authorrunning{M. Papoutsis et al.}

\institute{Department of Physics, University of Crete, 71003 Heraklion, Greece \email{mpapoutsis@physics.uoc.gr}
\and 
Institute of Astrophysics, FORTH, GR-71110 Heraklion, Greece
\and
Department of Physics and Institute of Theoretical and Computational Physics, University of Crete, 71003 Heraklion, Greece
\and
MIT Kavli Institute for Astrophysics and Space Research, Massachusetts Institute of Technology, Cambridge, MA 02139, USA
\and
Cahill Center for Astronomy and Astrophysics, California Institute of Technology, 1200 California Boulevard, Pasadena, CA 91125, USA
\and
Astronomical Institute of the Academy of Sciences, Boční II 1401, CZ-14100 Prague, Czech Republic
}


   \date{Received XXXX; accepted YYYY}

  \abstract
  {Recent intensive monitoring campaigns of active galactic nuclei (AGN) have provided simultaneous X-ray, UV, and optical data of unprecedented quality. The observations reveal a strong correlation between the UV and optical variability, but a weaker correlation between the X-ray and UV bands, challenging the standard X-ray reprocessing scenario. We revisit the X-ray/UV connection in NGC 5548 by fitting archival 2014 HST and \swift/XRT light curves assuming X-ray reverberation from a dynamically evolving X-ray corona. Our results show that, as long as the corona height, photon index and power vary over time, X-ray reverberation can explain the observed UV and optical variability within 2\% and 5\%, respectively (on average). The evolution of the best-fit parameters suggests that fast changes in coronal geometry and energetics on a time scale of days are required to explain the observed variability.} 
   

   \maketitle
%
\section{Introduction}

Active Galactic Nuclei (AGNs) exhibit variable X-ray and UV/optical emission across a range of timescales, with X-rays showing more rapid variability at short timescales. The physical mechanism responsible for the X-ray/UV/optical variability in AGN has been the focus of intensive study. \cite{Clavel92} were among the first to study the X-ray/UV variability in AGN by performing simultaneous X-ray and UV observations of NGC 5548. They observed correlated X-ray and UV variations, which they attributed to X-ray reverberation of the accretion disc. More recently, NGC 5548 was monitored by {\it Swift} over a two-year period \citep[from 03/2012 to 02/2014][]{McHardy14}. These observations showed a good overall correlation between the X-ray and UV/optical bands, with the UV/optical bands lagging the X-ray band with delays that increase with wavelength, consistent with expectations from X-ray reverberation. However, the measured lag amplitudes were found to be larger than expected.

Similar results were also reached by \cite{Edelson15}, \cite{Fausnaugh_2016} and \cite{Edelson19}, who studied the X-ray/UV and optical variability of NGC 5548 using the light curves from the AGN STORM campaign \citep{DeRosa15}. This was a dense and long monitoring campaign, involving {\it Swift}, HST and many ground-based telescopes. Many more monitoring campaigns of Seyferts have been conducted since then (see \cite{Cackett21} and \cite{Paolillo25} for a recent review of these campaigns). The NGC 5548 observations revealed strong and well-correlated UV/optical variability, yet again with time lags larger than predicted by standard disc reprocessing. \cite{Neustad202} suggested that temperature fluctuations propagating slowly through the accretion flow on timescales far longer than the light-travel time could explain these observations.

Since then, alternative ideas have been proposed to account for the larger than expected UV/optical time lags. For example, \cite{Netzer22} suggested that the observed time lags in NGC 5548 (and other Seyferts) are due to the response of the diffuse gas in the broad line region to the variable ionising continuum. \cite{cai20}  attributed the observed time lags to disc turbulences when the effect of large-scale turbulence is considered. 
Furthermore, \cite{Hagen24} suggested that the observed time-lags in Fairall 9 are due to the  reverberation of the variable EUV from an inner wind, which produces a lagged bound-free
continuum that matches the observed UV/optical lags. 

Despite the initial suggestions, X-ray reverberation in the lamp-post geometry can explain the UV/optical time delays \citep{Kammoun212, Kammoun23}, the power spectrum in the UV/optical bands \citep{Panagiotou22a}, and the frequency-resolved time lags \citep{Panagiotou25} of NGC 5548, using the STORM light curves. The same model can also fit the mean UV/optical/X-ray spectral energy distribution (SED) of the source \cite{Dovciak22}. These results suggest that, at least in NGC 5548, X-ray reverberation can explain the observed UV/optical variations.

However, there is still an issue that may suggest otherwise. This is because the X-ray to UV correlation is weaker than the UV to optical correlation \citep{Edelson15,Edelson19}. In fact, visually, the X-ray light curve does not appear to be the driver of the UV/optical light curves. \citet{Starkey17} showed that the inferred driving light curve, reconstructed from UV/optical data, did not match the observed X-ray light curve, while \citet{Gardner17} found that X-ray reverberation of a standard accretion disc results in simulated UV/optical light curves that reproduce too much of the hard X-ray high-frequency power. More recently, \cite{Mahmoud20}, \cite{Mahmoud23} and \cite{Hagen23} showed similar discrepancies between the X-ray reverberating disc signal and the observed UV light curves in NGC 4151, Ark 120 and Fairall 9, respectively. They proposed that intrinsic fluctuations within the accretion disc, rather than direct X-ray reprocessing, dominate the observed UV variability.
 
\cite{Panagiotou22b} showed that a dynamic corona can explain the possibility of a low cross-correlation between the X-ray and UV/optical light curves in the case of X-ray reverberation. \citet{Kammoun24} (K24 hereafter) further demonstrated that the time-variable SEDs of NGC\;5548 can be well reproduced by X-ray reverberation in the case of a dynamic corona, even in the presence of a weak X-ray/UV flux correlation. 

Here, we adopt a complementary approach by fitting the full light curves of NGC\;5548, which allows us to explore the temporal evolution of the disc–corona system in more detail. In particular, we show that the observed UV (and optical) light curves of NGC\,5548 during the STORM campaign can be naturally explained by reprocessing of X-rays emitted from a dynamical X-ray corona in the lamp-post geometry, where the corona power, height, and photon index are variable. Section \ref{sec:rev_model} introduces our reverberation model, followed by a description of the observational data in Section \ref{sec:data}. The fitting procedure is detailed in Section \ref{sec:fitting}, and the results are presented in Section \ref{sec:results}. Finally, we discuss our findings in Section \ref{sec:discussion}.

\section{The X-ray reverberation model}
\label{sec:rev_model}

We consider an X-ray source that illuminates an accretion disc. Part of the X-rays falling on the disc will be re-emitted in X-rays, while the other part will be absorbed by the disc and act as an additional source of heating. In this way, the disc emission is connected with the X-ray emission. In the case of a steady-state system, this connection can be expressed as:
\begin{equation}
    F_{\rm \lambda}(t) = F_{\rm \lambda, NT} +
     \int_{0}^{\infty} \Psi_{\lambda}(t')L_{X}(t-t')dt' \;,
\label{eq:discflux}
\end{equation}
where $F_{\rm \lambda}(t)$ is the total flux of the disc at wavelength $\lambda$ and time $t$, $F_{\rm \lambda, NT}$ is the constant flux of a standard accretion disc when there is no illumination \citep{NT73}, $L_{X}(t)$ is the luminosity of the X-rays at time $t$, and $\Psi_{\lambda}(t)$ is the disc response function, which describes how the disc responds to the illumination of X-rays \citep[see][]{Kammoun211, Kammoun23}. The shape and amplitude of the response function depend on 
all the physical parameters of the accretion disc/X-ray system, such as the black hole (BH) mass, the accretion rate, the height,  and the luminosity of the X-ray source, among others. These parameters may not be constant over time, and, as a result, the response function will evolve over time. In such a dynamic system, the disk emission may be estimated by generalising the above equation \citep{Panagiotou22b}
\begin{equation}
    F_{\rm \lambda}(t) = F_{\rm \lambda, NT}(t) + \sum_{i=0}^{N-1}
     \int_{t_i}^{t_{i+1}} \Psi_{\lambda, i}(t')L_{X}(t-t')dt' \;,
\label{eq:discflux2}
\end{equation}
where $t_0 = 0$, and $N$ is the number of times that $\Psi(t)$ changes. In this equation, we assume that $\Psi(t)$ remains constant within the time interval $(t_i, t_{i+1})$, while N can be arbitrarily large.

\section{The light curves}
\label{sec:data} 

In this work, we are interested in the UV and X-ray light curves of NGC\,5548. We first present the HST UV continuum light curves, followed by the X-ray light curve derived from \swift/XRT observations.

\subsection{The UV light curves}
For the UV, we use the HST H1 ($\lambda=1158$\AA), H3 ($\lambda=1479$\AA), and H4 ($\lambda=1746$\AA) continuum light curves of \cite{Fausnaugh_2016} (i.e., the shortest, the middle, and the longest wavelength HST light curves of the STORM campaign).
We consider only the far UV light curves because the convolution integral in Eq.(\ref{eq:discflux2}) can be computed fast in this case, since the width of the disc response is small ($\sim 1$ day) at these short wavelengths. In addition, they are not contaminated by the host galaxy light and, by construction, should not include any contribution by emission lines. 

We corrected the light curves for Galactic extinction with $E(B-V) = 0.0171$ \citep{Schlafly_2011} and the extinction law of \citet{Cardelli89}. We also corrected for absorption from the host galaxy, assuming $E(B-V)_{\rm host} = 0.14$, as determined by K24, and the extinction law of \citet{Czerny04}. The open grey squares in panels (b), (c), and (d) of Fig.\,\ref{fig:lightcurves} show the resulting H1, H3, and H4 band light curves, respectively. 

\subsection{The X-ray light curve}
\label{sec:xray}
 \begin{figure}
   \centering
    \includegraphics[width=9cm, height=7cm]{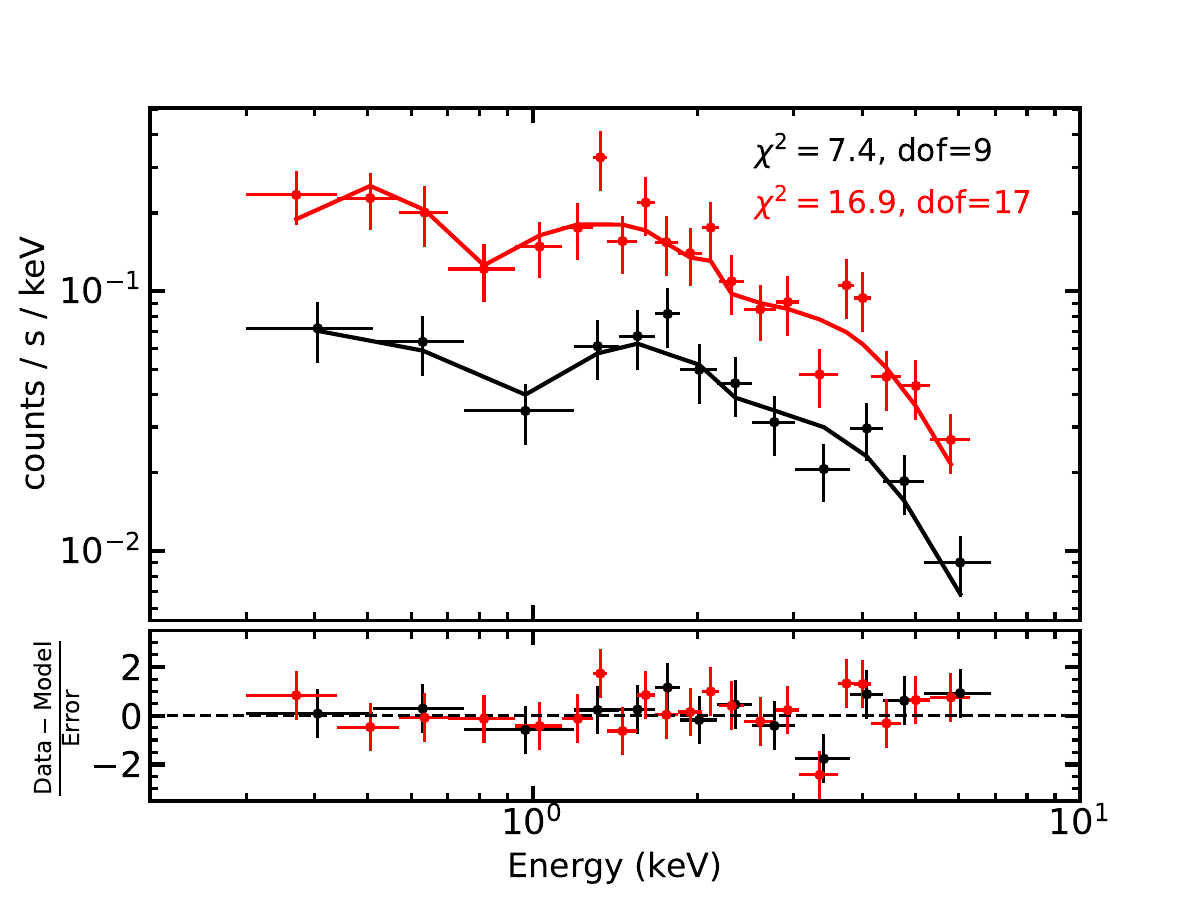}
      \caption{The \swift/XRT spectra for an observation with low count rate (black points) and an observation with high count rate (red points). The black and red lines show the best-fitting model of Eq.(\ref{eq:xspec}). The bottom panel shows the best-fit residuals.}
\label{fig:xrt}
\end{figure}

The X-ray luminosity in Eq.(\ref{eq:discflux2}) refers to the X-ray luminosity in the 2--10 keV band. To compute it, we calculated the flux in the 2--10 keV band by fitting the X-ray spectrum of each observation. We used the automatic \swift/XRT data products generator \footnote[1]{\url{https://www.swift.ac.uk/user_objects/}} \citep{Evans09} to extract the X–ray spectrum of each observation. We grouped each spectrum to have at least 15 counts per bin and fit the spectra in \texttt{XSPEC} \citep{Arnaud96} with an absorbed power law model of the form,
\begin{equation}
\textrm{model} = \texttt{TBabs} \times \texttt{zTBabs} \times \texttt{zxipcf} \times \texttt{cflux} \times \texttt{powerlaw}\;.
\label{eq:xspec}
\end{equation}
The \texttt{TBabs} component \citep{Wilms00} quantifies the X-ray absorption by our own Galaxy, and we fixed the column density to $N_{\mathrm{H}} = 1.55 \times 10^{20} \mathrm{cm}^{-2}$ \citep{HI4PI}. The second neutral absorber, \texttt{zTBabs}, accounts for absorption from the host galaxy, and we fixed its column density to $N_{\mathrm{H,host}} = 8.3 \times 10^{21} E(B{-}V)_{\mathrm{host}} = 12 \times 10^{20}\mathrm{cm}^{-2}$ \citep{Liszt21}. We also included a warm absorber using \texttt{zxipcf} \citep{Reeves08}. Since NGC 5548 exhibits variable ionised obscuration in X-rays \citep[e.g.][]{Mehdipour16, Dehghanian19_storm}, we allowed both the column density, $N_{\rm wa}$, and the ionisation parameter of the warm absorber, log($\xi$), to vary freely during the fit. Given the quality of the individual spectra, we cannot allow the covering fraction to also be variable; therefore, we fixed it to unity. 

Figure\;\ref{fig:xrt} shows the \swift/XRT spectrum of two observations: one with count rate of 0.2 (black points, ID 00030022237; exposure 964 s) and one with count rate of 0.6 (red points, ID 00033204104; exposure 570 s), which are representative of the low and high count rate spectra of the campaign. The best-fit models (defined by Eq.(\ref{eq:xspec})) are shown with the black and red lines, while the best-fit residuals are shown in the bottom panel. The best-fit parameters for the low and high count rate spectra are: $N_{{\rm wa},l} = 4.2^{+2.5}_{-1.8} \times 10^{22}$cm$^{-2}$, log$(\xi_l) = 1.4^{+0.6}_{-0.2}$, $\Gamma_l = 2.2^{+0.5}_{-0.6}$, and $N_{{\rm wa},h} = 4.6^{+2.8}_{-1.8} \times 10^{22}$cm$^{-2}$, log$(\xi_h) = 1.9^{+0.2}_{-0.6}$, and $\Gamma_{h} = 2^{+0.5}_{-0.2}$. The errors were estimated using the \texttt{error} command in \texttt{XSPEC} for $\Delta\chi^2$=2.71, which corresponds to the 90\% confidence range for a single parameter of interest. They can be large, but the main purpose of fitting the individual spectra is to compute the 2--10 keV unabsorbed X-ray flux (and then luminosity) of the source. 

\begin{figure*}
   \centering
    \includegraphics[width=17.5cm, height=13.5cm]{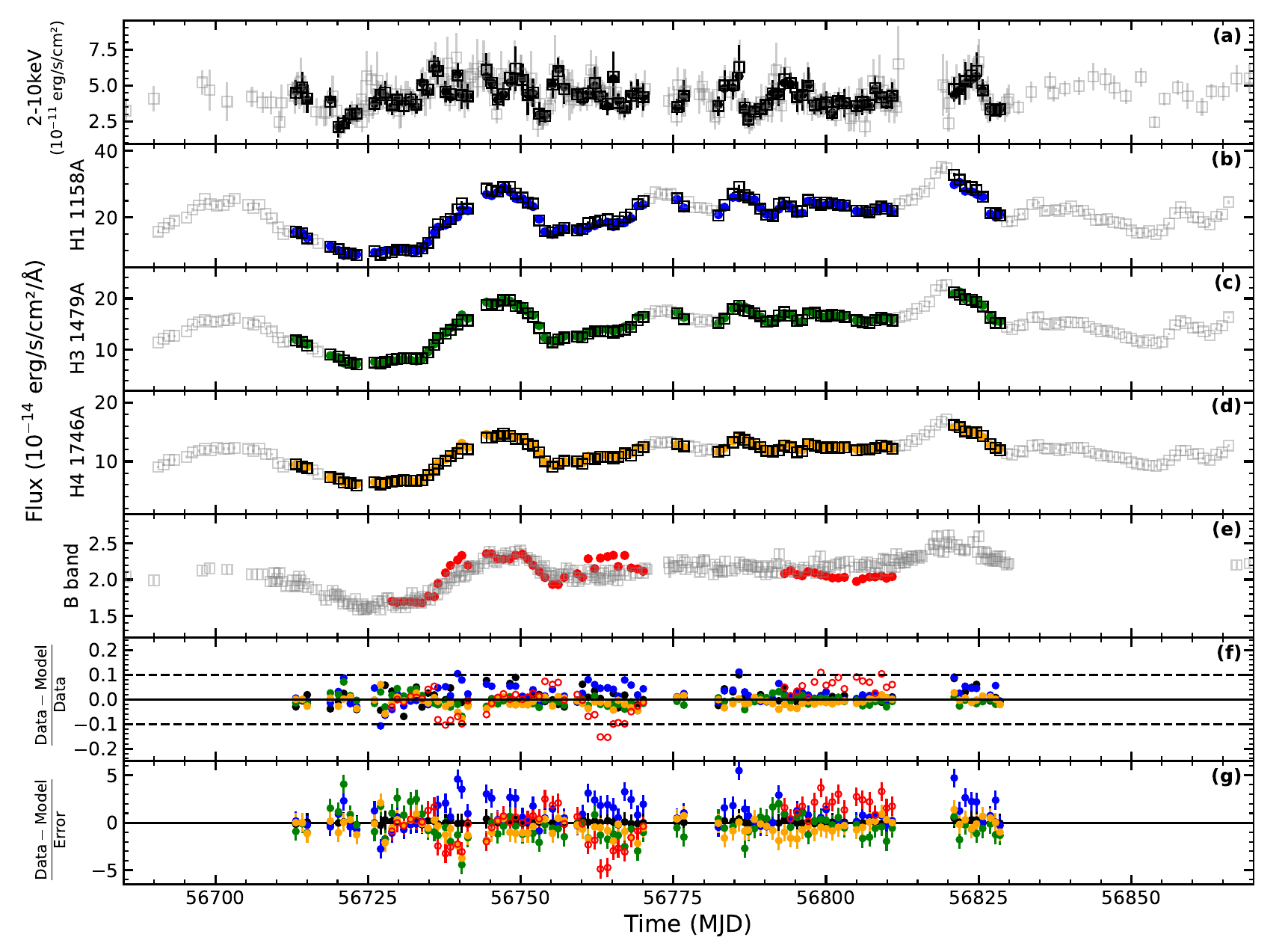}
      \caption{The grey open squares in panels (a), (b), (c), (d), and (e) show the observed 2--10 keV, H1, H3, H4 and B-band light curves, respectively. The black open squares and the filled coloured circles show the observations we considered in the fitting procedure and the best-fit model predictions, respectively. Panels (f) and (g) show the (data-model)/data and (data-model)/error ratios for the X-ray, the H1, the H3, and H4, and the B-band light curves (black, blue, green, orange, and red circles, respectively).}
         \label{fig:lightcurves}
   \end{figure*}

We use the \texttt{cflux} component to compute the 2–10 keV flux, $F^{\rm obs}_{2-10}$, and its error, $\sigma^{\rm obs}_{2-10}(t)$ (estimated as explained above for the best-fit parameters~). 
The unabsorbed flux calculated from the best-fit model to the spectra plotted in Fig.\;\ref{fig:xrt}  is $6.3 \pm 1.8 \times 10^{-11}$ and $4.6 ^{+0.9}_{-1} \times 10^{-11}$ erg/s/cm$^2$ for the low and high count rate spectra, respectively. The grey squares in the top panel of Fig.\,\ref{fig:lightcurves} show the resulting $F^{\rm obs}_{2-10}$ as a function of time for the whole monitoring campaign. We note that, in some cases, the flux error was quite large due to the poor quality of the single-observation spectra. For this reason, we omitted the points with $F^{\rm obs}_{2-10}/\sigma^{\rm obs}_{2-10} < 2$ ($\sim 10\%$ of the total points; these points are not shown in Fig.\,\ref{fig:lightcurves}).

In addition to the statistical error of the flux measurements, systematic uncertainties may also be present if the X-ray absorption is not modelled properly.
To test this, we assumed that the warm absorber is not variable and recalculated the unabsorbed flux of each observation using the best-fit values reported in Table 2 of K24 that were estimated when fitting the mean spectrum of the source ($N_{\rm wa} = 2.54 \times 10^{22}$cm$^{-2}$, log$(\xi) = 1.37$). We find that the $F^{\rm obs}_{2-10}$ values do not vary by more than $\sim 25$\%. This difference is almost always smaller than the error of the flux measurements. 

The X-ray flux measurements include contributions from both the primary emission and the X-ray reflected component from the accretion disc. However, we only need the primary emission for our modelling. We attempted to unravel the primary and reflected components by modelling the reflection with \texttt{xillver} \citep{garcia_16} and refitting the spectra. However, because of the low statistical quality of individual spectra, the reflected flux could not be constrained. We therefore assumed that 10\% of the observed X-ray flux arises from reflection and 90\% from the primary emission. This assumption is supported by the spectral modelling of the mean SED of NGC 5548 using \texttt{KYNSED} \citep{Dovciak22} in K24, which indicates that the reflected component contributes around 10\% of the observed 2–10 keV flux. We further checked that our best-fit results are consistent with this assumption a posteriori using \texttt{KYNSED}, as explained in Sect.~\ref{sec:results}.

\section{The X-ray reverberation modelling of the UV light curves}
\label{sec:fitting} 

We use the relativistic reverberation model \texttt{KYNXiltr} \citep{Kammoun23} to compute \psilt. In summary, a lamp post geometry is assumed for the X-ray source, the emission of which follows a power law with a slope of $\Gamma_{\rm int}$ and a high energy exponential roll-off, which we fix at $E_{\rm cut}=150$ keV \citep{Ursini15}. We fix the mass of the central BH at $M_{\rm BH}=7\times 10^7$ M$_{\odot}$ \citep{Horne21}, the inclination of the disc at $\theta=40^o$ \citep{Pancoast14}, and the distance at $D_{L}=80.1$ Mpc according to the NASA/IPAC Extragalactic Database (NED). For the BH spin, the colour correction factor, and the accretion rate (in Eddington units), we accept \spin=0, \fcol=1.7 (the best-fit values in K24), and \medd=0.06. We also tested the cases of \medd=0.05 and 0.07, and found that a value of 0.06 provides a better fit to the UV light curves. The free parameters are the height of the X-ray source, $h$, the photon index $\Gamma_{\rm int}$, and the ratio \ltransf, where $L_{\rm disc}$ is the power released by the accretion process in the disc and $L_{\rm transf}$ is the luminosity of the X-ray corona which we assume is equal to the accretion power released in the disc below a certain radius $r_{\rm transf}$ \citep{Kammoun23}.

We note that the photon index of the corona emission, $\Gamma_{\rm int}$, is left as a free parameter. This is because the best-fit photon index from the spectral modelling of the individual energy spectra, $\Gamma_{\rm obs}$, cannot be used in our modelling for various reasons. First of all, the X-ray and UV observations are not strictly simultaneous. Secondly, the uncertainty of the $\Gamma_{\rm obs}$ values can be quite large, as discusses in the previous section.  In addition, $\Gamma_{\rm obs}$ was estimated from a model that did not include reflection from the disc (and, to a lesser extent, from a distant torus as well). As a result, $\Gamma_{\rm obs}$ may be different from the intrinsic slope of the X-ray spectrum, $\Gamma_{\rm int}$. Therefore, we need to let $\Gamma_{\rm int}$ free during the modelling of the UV light curves. However, although we cannot directly use $\Gamma_{\rm obs}$ to constrain $\Gamma_{\rm int}$, we can use $F^{\rm obs}_{2-10}$ to constrain it (assuming that, even if we fit the observed spectra with phenomenological models, the unabsorbed flux we compute is indicative of the intrinsic flux). This can be done using the \texttt{KYNXiltr} code to compute the model flux in the 2--10 keV band, $F^{\rm mod}_{2-10}$, which depends on $\Gamma_{\rm int}$, and then fit the unabsorbed $F^{\rm obs}_{2-10}$ light curve simultaneously with the UV light curves.

We computed the disc response function using a uniformly sampled grid of model parameters, with 20 linearly spaced values for each of the free parameters: coronal height $h$ ranging from 5 to 50 $r_g$, \ltransf\ from 0.3 to 0.9, and photon index $\Gamma_{\rm int}$ from 1.3 to 2.1. These parameter ranges are motivated by the best-fit values of K24. To compute $F_{\lambda}(t)$ using Eq.\,(\ref{eq:discflux2}) we also need the disc flux when the disc is not illuminated by the X-rays, $F_{\rm \lambda, NT}(t)$. We calculate this using \texttt{KYNSED}, with the same model parameters as above.
Note that $F_{\rm \lambda, NT}(t)$ can also be variable if \ltransf\ is variable. The last parameter that we need is $L_{X}$. As we mentioned previously, $L_X$ in this equation refers to the 2--10 keV band X-ray luminosity. This is equal to 0.9$\times$\fobst$\times4\pi D_L^2$, where the factor of 0.9 accounts for the fact that we only need the primary emission, as discussed in the previous section.

We cannot reliably predict all the observed UV flux measurements using the convolution integral of Eq.\,(\ref{eq:discflux2}) because this requires well-sampled X-ray observations (roughly two per day) over a period of $\sim$ 3 days preceding each UV data point. The black boxes in panels (b), (c), and (d) of Fig.\ref{fig:lightcurves} indicate the HST measurements that we fit (91 points in each light curve). We start from MJD 56713 because the first $\sim 25$ days of the X-ray light curve are too sparsely sampled. This is also the case with the last $\sim 40$ days of the X-ray light curve, whereas the gaps in the X-ray light curve around MJD $56773$, $56780$, and $56815$ also prevent the model prediction of the UV flux at these times.  

For each set of model parameter values, we computed the model fluxes in the H1, H3, and H4 bands, using Eq.\,(\ref{eq:discflux2}). Since the width of the response functions in these bands should be small, we stop the calculation of the convolution integral in Eq.\,(\ref{eq:discflux2}) at $t'=3$ days. This choice significantly increases the computational efficiency of the model fitting process and does not compromise the accuracy of the model estimates (we verified that contributions from longer times are below $\sim 1\%$). 

For the first HST measurements at $t=56713$ MJD, the convolution reduces to the simpler expression in Eq.\,(\ref{eq:discflux}). The best-fit model parameters at this time are determined by minimising the weighted squared error (WSE) between the model and the observed H1, H3, H4 and $F_{2-10}^{\rm obs}$ fluxes,

\begin{equation}
    \textrm{WSE}(t) = \left( \frac{F^{\rm mod}_{2-10}(t) - F^{\rm obs}_{2-10}(t)}{F^{\rm obs}_{2-10}(t)} \right)^2 +\ \sum_{i=1}^3 \left( \frac{F^{\rm mod}_{\lambda,i}(t) - F^{\rm obs}_{\lambda,i}(t)}{F^{\rm obs}_{\lambda,i}(t)} \right)^2.
    \label{eq:fit}
\end{equation}

We adopt the WSE approach as it enables a comparison between model and data in terms of fractional differences, without relying on potentially unreliable or underestimated observational uncertainties. This method also ensures that the X-ray light curve contributes to the overall fit as much as a UV light curve, despite having larger error bars.

As we explained above, we fit $F^{\rm obs}_{2-10}$ together with the UV light curves, mainly to constrain $\Gamma_{\rm int}$. We note that because the \swift\ and HST observations are not simultaneous, we used linear intrerpolation to compute \fobst\ at each time of the UV light curve points. The black boxes in the top panel of Fig.\,\ref{fig:lightcurves} show the interpolated \fobst\ values. Once the best-fit parameters (that is, $h$, \ltransf, $\Gamma_{\rm int}$) are obtained for the first time point, we proceed iteratively, using Eq.\;(\ref{eq:discflux2}) to fit the next light curve point by finding the parameters that produce the minimum WSE. In this way, the value of the model parameters can vary only at the times when the HST observations were made, but not between. 

 \begin{figure}
   \centering
    \includegraphics[width=9cm, height=9cm]{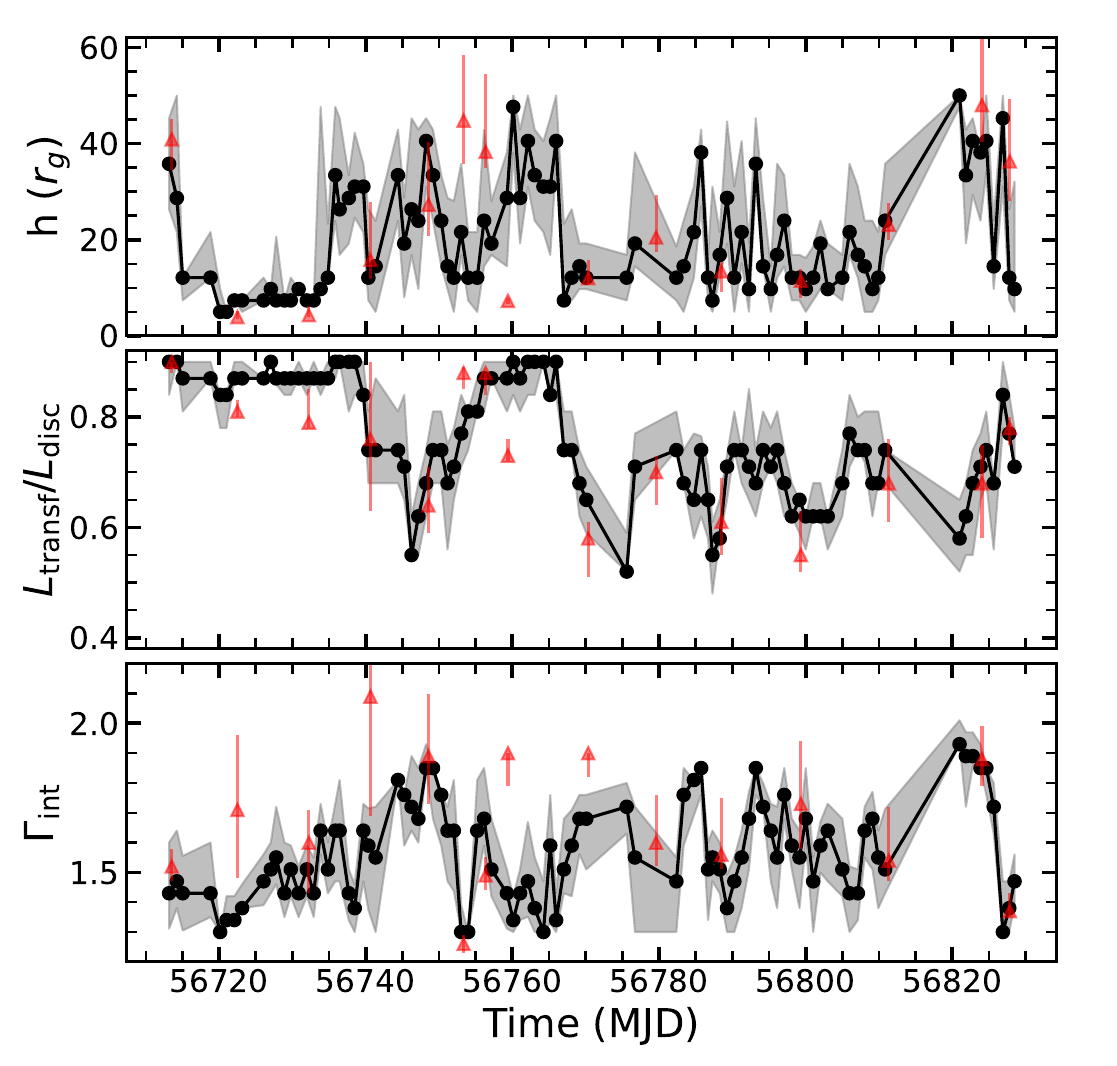}
      \caption{The best-fit parameters: height (top panel), \ltransf\ (middle panel), and the photon index (bottom panel). The grey shaded area shows the 1$\sigma$ confidence interval, which we computed by repeating the fitting procedure 100 times, each time resampling the X-ray light curve from Gaussian distributions centred on the measured values with standard deviations equal to their error bar. For computational efficiency, these fits were performed using a coarser parameter grid (i.e., 10 values per parameter instead of 20). The red triangle points show the results of K24.}
         \label{fig:parameters}
   \end{figure}

\section{Best-fit results}
\label{sec:results}

The black points in the top panel in Fig.\,\ref{fig:lightcurves} show the best fit 2--10 keV flux. The blue, green and orange points in panels (b), (c), and (d) show the best-fit H1, H3, and H4 light curves, respectively. 
Panel (f) of Fig.\,\ref{fig:lightcurves} shows the ratio (data-model)/data. The black dashed lines indicate discrepancies between the model and the data of the order of 10\%. We also show the best-fit residuals (i.e. (data-model)/error) in panel (g).

Figure\,\ref{fig:lightcurves} shows that X-ray reverberation is able to explain all the observed variability features in the UV light curves rather well, at all sampled time scales. The plot of the best-fit ratio residuals (panel (f) in Fig.\,\ref{fig:lightcurves}) shows that the model reproduces the X-ray and UV data with an accuracy better than 10\% throughout the monitoring period, capturing all variability features of the light curves. The average absolute residuals are 
$\sim$3\% for the H1 light curve and $\sim$2\% for the 2–10 keV, H3 and H4 light curves. This is indicative of the good agreement between the model and the observations. However, the best-fit $\chi^2$ is 517 for 91 degrees of freedom\footnote{As we explain in Sect.\;\ref{sec:fitting}, the model parameters are allowed to vary at each point in the observed light curve, giving in effect 3N free parameters, where N is the number of points in the light curve. Since we considered 4 light curves in the fit, the number of degrees of freedom is 4N-3N=N.}. From a statistical standpoint, this result indicates that the model does not fit the data well. The $\chi^2$ results indicate that the unaccounted 2-3\% of the observed variations (on average) is highly significant (assuming the photometric errors are determined correctly). If we add a systematic error of 4\% in quadrature, then the best-fit $\chi^2$ reduces to 110, which implies a good fit to the data with a null hypothesis probability of 0.09. We computed the excess variance of the H1, H3 and H4 light curves using only the points we modelled, and we found that their average variability amplitude (i.e. the square root of the excess variance) is 30\%, 25\% and 22\%, respectively. Therefore, we conclude that X-ray reverberation can explain well most of the observed variations in the UV light curves.  
The open squares and filled red circles in panel (e) of Fig.\,\ref{fig:lightcurves} show the \swift\ B-band flux measurements during the STORM campaign \citep{Edelson15} and the best-fit model predictions. We did not take into account the B-band data in our model fitting. This is because the disc response function in the B-band is much wider than at the UV wavelengths, so the convolution integral in Eq.\,(\ref{eq:discflux2}) must be computed up to (at least) $t' \sim 10$ days. This considerably increases the computational time required to model the B-band light curve. In addition, there are fewer time intervals of well-sampled X-ray observations (i.e. at least two X-ray observations per day over a period of 10 days) to predict the B-band flux. For this reason, we plot the best-fit model B-band fluxes only over the MJD $\sim 56729 - 56770$ and $\sim 56793 - 56811$ periods in Fig.\,\ref{fig:lightcurves}.

Nevertheless, the agreement between the model and the observed B-band light curve is also quite good (the average absolute residual is 5\% in this case), although there are some discrepancies. For example, the model systematically underestimates the flux observed between $\sim 56800-56809$ MJD. This could be due to many factors. 
First, we have not taken into account the (variable) contribution of the emission lines, which can be of the order of $\sim 10\%$ in this band \citep[see e.g. Table 8 in][]{Fausnaugh_2016}. Secondly, the model assumes a flat disc, which is an oversimplification since the disc height should increase with radius. Even small disc heights can affect the reverberation flux in the optical bands, depending on the value of the model parameters.  

\subsection{Best-fit parameters}

Figure\,\ref{fig:parameters} shows the evolution of the best-fit parameters with time.\footnote{ We checked with \texttt{KYNSED} that the X-ray reflection flux in the 2--10 keV band that the model predicts is $\sim 10$\% of the total flux that a distant observer will measure in all cases (i.e, for all the best-fit parameter combinations plotted in Fig.\,\ref{fig:parameters}). This implies that the best model fit is consistent with our original assumption regarding the fraction of the X-ray reflection flux in the observed spectra.} The top panel shows the height of the X-ray source, the middle panel shows the energy transfer parameter \ltransf, and the bottom panel shows the best-fit photon index $\Gamma_{\rm int}$. All parameters need to be highly variable to reproduce the observed X-ray and UV variability. The red triangles show the best-fit results of K24, who fitted time-resolved optical/UV/X-ray spectral energy distributions of NGC 5548, using data from the STORM campaign as well.

Our $\Gamma_{\rm int}$ measurements are very similar to the K24 measurements who fitted the SEDs using \texttt{KYNSED}, which includes the X-ray reflection component in the X-ray band. This result indicates that the $\Gamma_{\rm int}$ values we derive from the light curve fitting are consistent with those obtained from the spectral modelling. 
In general, the best-fit parameters from both approaches show very good agreement. For example, 10 (of the 15) of the best-fit $h$, \ltransf, and $\Gamma_{\rm int}$ values are consistent within 1$\sigma$. Only 4 of the (total) 45 best-fit values are outside the 2$\sigma$ range, and only one of the K24 best-fit values in the bottom panel of Fig.\,\ref{fig:parameters} shows a deviation larger than 3$\sigma$ from our results (the red point at MJD$\sim56760$ in Fig.\ref{fig:parameters}). The agreement between the best-fit results shows that the two approaches (i.e. fitting time-resolved broad-band SEDs and modelling the observed light curves) are consistent in tracing the evolution of the model parameters and their trends. 

\subsection{Correlations between the best-fit parameters}

The variability of the model parameters shown in Fig.\,\ref{fig:parameters} appears to be rather complicated, but some obvious trends are apparent. For example, the large-amplitude flare in the UV and optical band light curves between $\sim 56735-56755$ MJD is mainly due to a systematic variation of \ltransf\ in the same period. As \ltransf\ decreases, the disc emission in the UV and optical bands increases (because $r_{\rm transf}$ decreases), reaching a peak when \ltransf\ is minimum. Then the UV and optical disc flux decreases as \ltransf\ increases again.  However, it is not only the corona luminosity that drives the UV/optical emission. For example, the increase in the UV/optical flux observed from $\sim 56755$ to $\sim 56765$ MJD is mainly due to a variation in the height of the corona. During this period \ltransf\ remains roughly constant, but the height of the corona increases.  As a result, the X-ray flux that illuminates the disc also increases. 

\begin{figure}
   \centering
    \includegraphics[width=9cm, height=9cm]{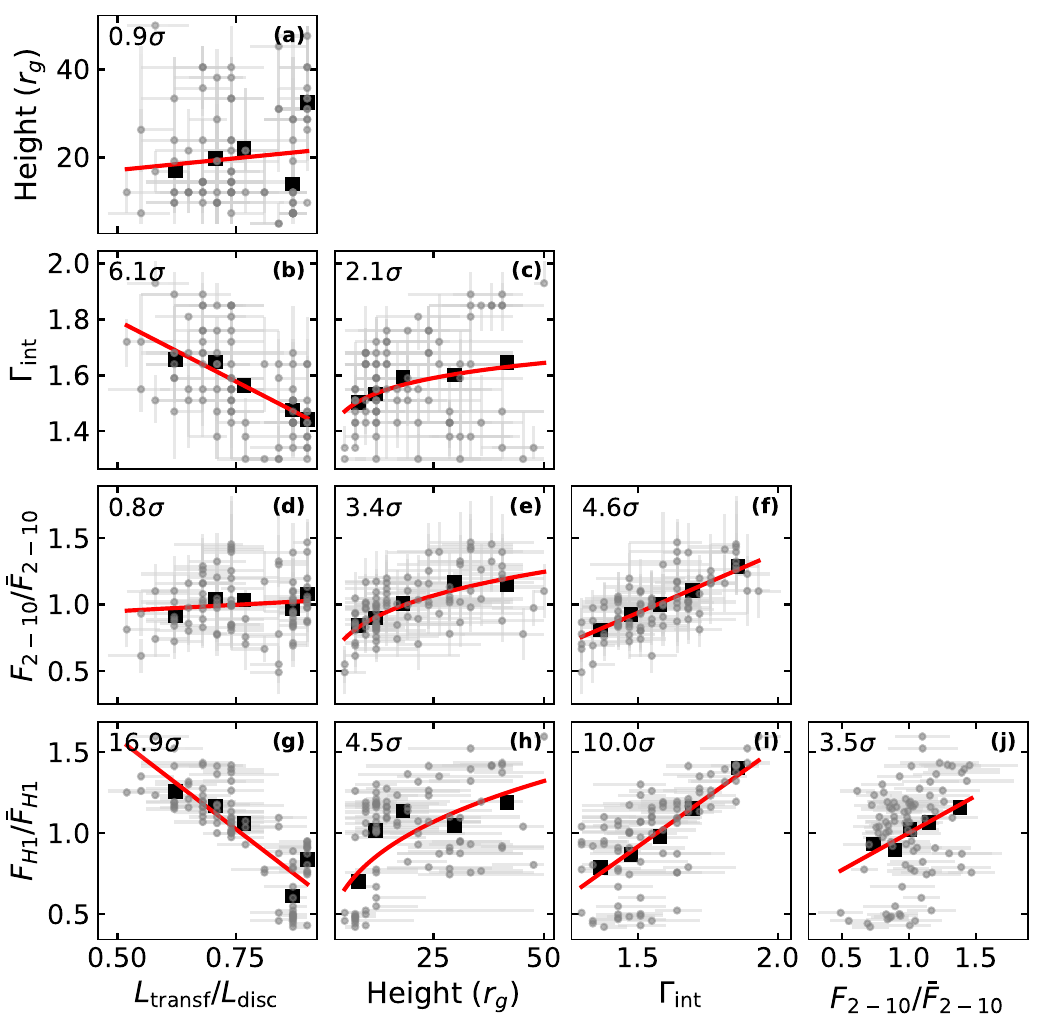}
      \caption{Correlations between the best-fit parameters i.e. the height, the photon index, and the energy transferred to the corona) and the X-ray (2--10keV) and UV observed fluxes. The red lines show the best-fit straight line or power law model, and the black squares show the binned data. The numbers in each plot show the significance that each slope is different from zero.}
\label{fig:cor}
\end{figure}

Figure\,\ref{fig:cor} shows the correlation between the various best-fit parameters and the observed X-ray and UV fluxes. We model these trends with a line in linear or logarithmic space using the ordinary least-squares (OLS) regression method \citep{Isobe90}. The best-fit lines are shown with the red solid lines in the various panels. In most cases, the regression provides a good description of the data. This is further illustrated by the binned averages (black squares; 20 points per bin), which follow the best-fit lines closely. 

Figure \ref{fig:cor} is similar to Fig. 8 in K24. Our work is closely related to their work, and both approaches recover similar parameter correlations. \cite{Kammoun24} constructed broad band SEDs using data from the STORM campaign and fitted them with \texttt{KYNSED}. They showed that X-ray reverberation can fit the time variable SEDs well, but were able to study only 15 spectra. Our work is performed in the time domain. We could fit only three UV light curves, but we used many more observations. Consequently, the panels in Fig. \ref{fig:cor} contain many more data points than those in Fig. 8 of K24, increasing the statistical significance of the detected correlations and providing tighter constraints on their evolution.

We find that many of the apparent correlations in Fig.\,\ref{fig:cor} are statistically significant\footnote{We consider a correlation significant when the best-fit slope is different from zero by at least $3\sigma$. We compute the error, $\sigma$, of the slope by resampling the parameters 1000 times from Gaussian distributions and performing the fit again. We take the standard deviation of the resulting best-fit slope distribution as the slope error.}. Panel (a) in Fig.\,\ref{fig:cor} shows that the corona height and total X-ray luminosity (i.e. the power transferred to the corona \ltransf) are not correlated. On the other hand, panel (b) of the same figure shows that the spectral slope, $\Gamma_{\rm int}$, is significantly anticorrelated with the total X-ray luminosity. An increase in X-ray luminosity probably increases the corona temperature (hence the flattening of $\Gamma_{\rm int}$) but does not affect the corona height. At the same time, the UV flux (i.e. $F_{H1}/ \bar{F}_{H1}$, where $\bar{F}_{H1}$ is the mean flux) strongly anticorrelates with the luminosity of the corona (panel g). This is because we have assumed that the corona is powered by the accretion process: when the fraction of the accretion power transferred to the corona decreases, the accretion power available for the heating of the disc increases.

 However, the 2--10 keV X-ray flux does not appear to be a good indicator of the total X-ray luminosity, as shown by the lack of a strong correlation between the observed $F_{\rm 2-10}$ (normalised to its mean, $\bar{F}_{\rm 2-10}$)  and \ltransf\ (panel (d) in Fig.\,\ref{fig:cor}). We suspect that this is due to two reasons: an increase in total X-ray luminosity should result in an increase in $F_{\rm 2-10}$. At the same time, the resulting flattening of $\Gamma_{\rm int}$ should also decrease $F_{\rm 2-10}$ (see panel f, which shows the `steeper when brighter' effect in Seyfert galaxies). The two effects probably cancel each other out, causing the lack of correlation between $F_{\rm 2-10}$ and the total X-ray luminosity. On the other hand, $F_{\rm 2-10}$ appears to correlate broadly with the height of the corona (panel (e) in Fig.\,\ref{fig:cor}). The correlation is moderate and appears to hold mainly at low heights (i.e. at heights lower than 15-20 r$_g$). In these cases, as the height increases,  more X-ray photons manage to escape to infinity, hence the increase in the number of the X-ray photons we observe. 

We note that since the UV flux and the X-ray luminosity are anticorrelated (panel (g) of Fig.\,\ref{fig:cor}), if the 2--10 keV flux were a good indicator of the total X-ray luminosity, then we would also expect the 2--10 keV and the UV flux to be anticorrelated (contrary to what is commonly expected). Panel (j) suggests a modest positive correlation between $F_{2-10}$ and UV flux, with a lot of scatter. This explains the small amplitude of the cross-correlation function between the X-ray and UV light curves (e.g. \cite{Edelson15,Fausnaugh16,Edelson19}). We believe that this rather weak and positive correlation is due to the fact that both $F_{2-10}$ and the UV flux are positively correlated with $\Gamma_{\rm int}$ (mainly), and the corona height. 

We have already discussed the correlation between $F_{2-10}$, $h$, and $\Gamma_{\rm int}$. Panels (h) and (i) show that the UV flux also correlates with corona height and $\Gamma_{\rm int}$. The first correlation is due to the fact that as the corona height increases, the corona subtends a larger angle to the disc, hence the amount of X-rays illuminating the disc (and getting absorbed) increases. The positive correlation between the UV flux and $\Gamma_{\rm int}$ is due to two reasons: $\Gamma_{\rm int}$ is correlated with the total X-ray luminosity, but, for a fixed X-ray flux, steeper $\Gamma_{\rm int}$  (slightly) increases the fraction of the flux that is due to X-ray absorption in the UV and optical bands (see Fig. 27 in \cite{Kammoun211}). 

\section{Discussion and conclusions}
\label{sec:discussion}

We present results from the modelling of the HST UV light curves and the \swift/XRT X-ray light curve of NGC\;5548 during the 2014 STROM campaign, assuming X-ray reverberation from a dynamical X-ray source. We find that X-ray reverberation within the context of the lamp-post model reproduces the observed light curves to within 2-3\% (on average), but only if the X-ray corona properties, that is the corona height $h$, the energy transferred from the disc to the corona, \ltransf, and the photon index, $\Gamma_{\rm int}$, are variable even on timescales of days. Although direct modelling of the optical light curves is more challenging, the X-ray reverberation model can also describe the B-band light curve well, with residuals being $\sim$5\%, on average. 

One possible physical explanation for the variations in corona parameters could be the ``failed jet" model of \citet{Ghisellini04}. According to this model, radio-quiet AGNs emit blobs of material which cannot reach escape velocity. Therefore, they reach a maximum radial distance and then fall back, colliding with blobs ejected later and still moving upward. 
These collisions dissipate the bulk kinetic energy of the blobs by heating the plasma, hence creating, in effect, X-ray-emitting regions, which naturally will have variable power, height, and photon index, in agreement with our results. 

In this model, the X-ray regions may have intrinsic velocity, and there is a possibility of multiple X-ray regions appearing simultaneously at different heights. Given these complexities, the success of our modelling of the observed UV (and optical) variations in NGC 5548 is rather remarkable. The inclusion of more than a single X-ray corona and/or the addition of the intrinsic velocity of the corona with respect to the accretion disc may account for the UV variability amplitude we cannot account for at the moment. 

These results, together with our previous results from the cross-correlation and power spectrum analysis of the STORM light curves, suggest that the X-ray reverberation model can account for most of the variability properties of NGC 5548. There are indications that, even if X-ray reverberation operates in some Seyferts, it may not be able to explain the full range of observed variations in AGN. For example, \cite{Beard25} found that X-ray reverberation cannot account for the long-term UV variations in NGC 4395. On the other hand, \cite{Papoutsis25} showed that X-ray reverberation could be responsible for the very low frequency variations of the SDSS quasars. 

A broader examination of mean SEDs, power spectra, and interband time lags across a larger sample of sources is needed as a further test of the X-ray reverberation framework. Similarly, modelling the UV and optical light curves of additional Seyfert galaxies could help assess whether X-ray reverberation from a dynamical corona is capable of producing the observed UV/optical variability, despite the relatively weak X-ray/UV and optical correlations. Such detailed studies of as many sources as possible can show whether X-ray reverberation can explain (at least most of) the observed variations.

\begin{acknowledgements}
We thank the referee for their useful comments that helped us improve the manuscript.
\end{acknowledgements}

%
%

\bibliographystyle{aa}
\bibliography{main}

\end{document}